\documentclass[namedreferences,journalpaper,10pt,final,twoside,onecolumn]{SolarPhysicsMod}
\usepackage[optionalrh,solaenum]{spr-sola-addons} 
\usepackage{graphicx}                    
\usepackage{color}                       
\usepackage{url}                         
\usepackage{cases}
\usepackage{marvosym}

\usepackage{mathptmx}

\usepackage{solaheader}	
\renewcommand{\solayear}{2010}

\begin{document}

\begin{article}

\begin{opening}

\title{Diverging and Converging Flows around Sunspot Structures in Rotating and Non-Rotating Axisymmetric MHD Simulations}

\author{T.~Hartlep \sep F.H.~Busse \sep N.E.~Hulburt \sep A.G.~Kosovichev}

%
\runningauthor{Hartlep {\sl et~al.}}
\runningtitle{Diverging and Converging Flows around Sunspot Structures}

 \institute{
  T.~Hartlep (\,{\large\Letter}) \sep A.G.~Kosovichev \\
  W.W.~Hansen Experimental Physics Laboratory, Stanford University, Stanford, CA, USA \\
  email: \url{thartlep@sun.stanford.edu} \\
  \vspace*{1em}
  F.H.~Busse\\
  Institute of Physics, University of Bayreuth, Bayreuth, Germany \\
  \vspace*{1em}
  N.E.~Hulburt \\
  Lockheed Martin Solar and Astrophysics Laboratory, Palo Alto, CA, USA
  }

\begin{abstract}

We present results on modeling solar pores and sunspots using 2D axisymmetric magneto-hydrodynamic (MHD) simulations.
These models are helpful for understanding the mechanisms of magnetic field concentration in sunspots, and the large-scale flow patterns associated with them.
The simulations provide consistent, self-maintained, although not fully realistic, models of concentrated magnetic field near the solar surface.
In this paper, we explore under which conditions the associated flows are converging or diverging near the surface.
We find that in most cases in which a stable, pore-like concentration of magnetic field forms, a configuration with converging over diverging flow is established. 

\end{abstract}

%

\end{opening}


\section{Motivation and Objectives}

The mechanisms of how magnetic pores and sunspots form on the Sun are still poorly understood. 
Observations and numerical simulations suggest that their structure is intimately linked with characteristic surface and subsurface flows in and around the magnetic field concentration.
Fully developed sunspots exhibit a surface outflow in their penumbra, the so-called Evershed flow, presumably caused by interaction between the near-surface granular convection and the highly inclined penumbral magnetic field as suggested by the numerical simulations of \inlinecite{2009ApJ...700L.178K}.
Observation ({\sl e.g.}, \opencite{2010ApJ...708..304Z}) have revealed downflows in the central region of the sunspot and subsurface converging flows (inflows) below the granulation layer, as well as outflows further below.
Numerical models of subsurface magneto-convection in axisymmetric configuration \cite{2000MNRAS.314..793H,2006MNRAS.369.1611B,2008MNRAS.387.1445B} have been able to reproduce similar flow structures.
In most of their simulations, inflows over outflows were found.
It is conjectured that this flow is fundamentally important for maintaining the integrity of the magnetic field concentration.
Recently, a numerical simulation of realistic compressible magneto-hydrodynamic (MHD) convection has been reported in which a magnetic field concentration was held in place by boundary conditions \cite{2009Sci...325..171R}.
A realistic Evershed flow was found at the surface, but the outflow continued well below the surface.  
\opencite{2006MNRAS.369.1611B} and \opencite{2008MNRAS.387.1445B} did in fact also found cases with a diverging flow over a converging flow.
The present paper extends their work exploring in more detail under which conditions diverging or converging flows can hold a magnetic flux concentration in place.

\section{Numerical method}

We study magneto-convection in an axisymmetric cylindrical geometry using a code originally developed by \inlinecite{2000MNRAS.314..793H} for the two-dimensional (2D) case and later extended by \inlinecite{2008MNRAS.387.1445B}.
The exact equations and their parameters are discussed in detail in these two publications.
The model considers a layer of electrically conducting, perfect monoatomic gas subject to uniform gravitational acceleration, with constant shear viscosity, magnetic diffusivity, and magnetic permeability.
It approximates the conditions in the upper part of the solar convection zone but excluding the very top few hundred kilometers below the photosphere where the plasma is only partially ionized and radiation needs to be accurately modeled. 
For thermal boundary conditions, we prescribe a constant heat flux at the bottom and StefanÕs law at the top.
The side wall is perfectly electrically conducting and does not allow for a heat flux across it. 
Top, bottom, and outside walls are impenetrable and stress free. 
The magnetic field is vertical at the bottom and matched to a potential field at the top.

The computational domain can be subjected to constant rotation by including Coriolis and centrifugal forces.
The equations are written and solved in non-dimensional form with the key parameters being: Rayleigh number, $R$,  Prandtl number, $\sigma$, magnetic Prandtl number, $\sigma_m$, temperature contrast between the top and bottom of the domain, $\theta$, rotation rate, $\Omega$, aspect ratio, the ratio between height and radius of the cylindrical domain, $\Gamma$, and the Chandrasekhar number, $Q$, a measure of the magnetic flux in the system.
The ratio between specific heats is chosen to be $\gamma=5/3$, appropriate for a monoatomic ideal gas.

The initial temperature and density profiles in the simulations take the form of a polytrope, in non-dimensional form $T(z) = 1 + \theta z, \rho(z)=(1+\theta z)^m$, where $T$, $\rho$, $z$, and $m$ are the non-dimensional temperature, density, depth (ranging from 0 at the top of the domain to 1 at the bottom), and the polytropic index, respectively.
Simulations are started with an initial uniform vertical magnetic field and are run until the system reaches a more or less steady state, if such a state is attained at all for the given set of parameters. 
The results presented in this paper are all for an aspect ratio of $\Gamma=3$.

The equations are solved using a finite-difference scheme accurate to sixth order and a fourth-order time marching scheme.

\section{Results}

\begin{figure}
	\centering
	\vspace*{0.25cm}
	\includegraphics[width=5.6cm]{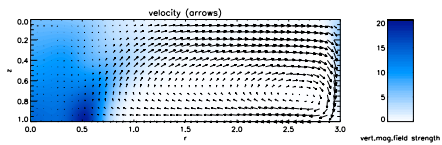}
	\vspace*{0.25cm}
	\caption{Flow visualization (arrows) and strength of the vertical magnetic field (contours) for a case that forms a diverging flow over a converging flow. See text for parameters.}
	\label{Fig:Results:Repro}
\end{figure}

\begin{figure}
	\centering
	\vspace*{0.25cm}
	\includegraphics[width=5.6cm]{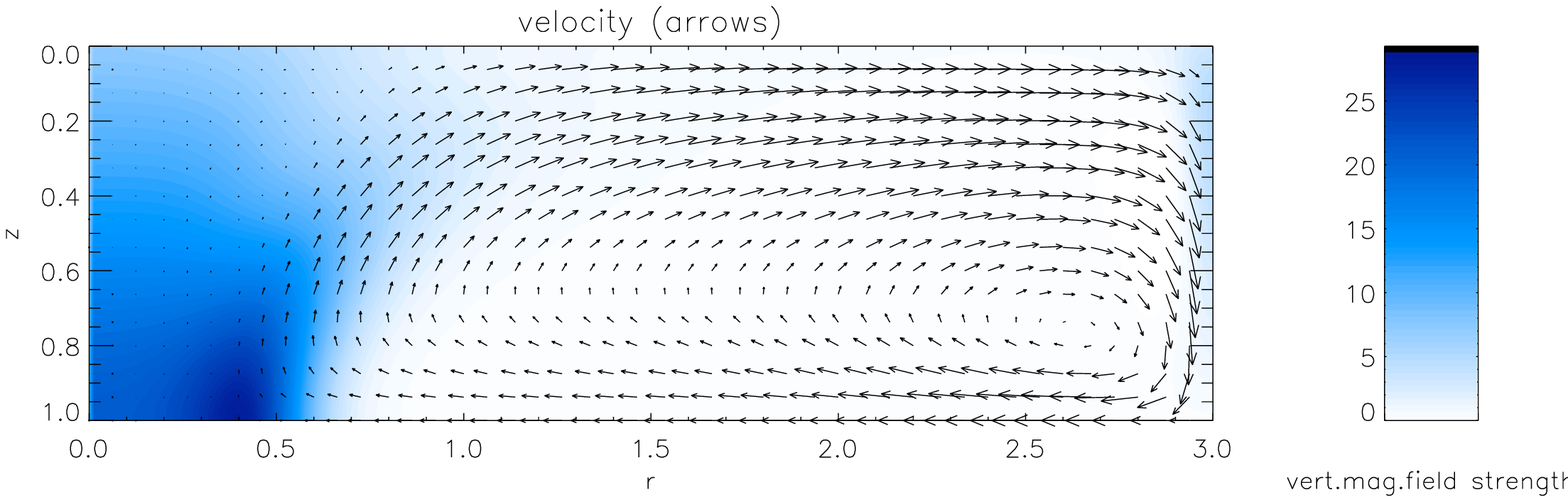} \hspace*{0.5cm}
	\includegraphics[width=5.6cm]{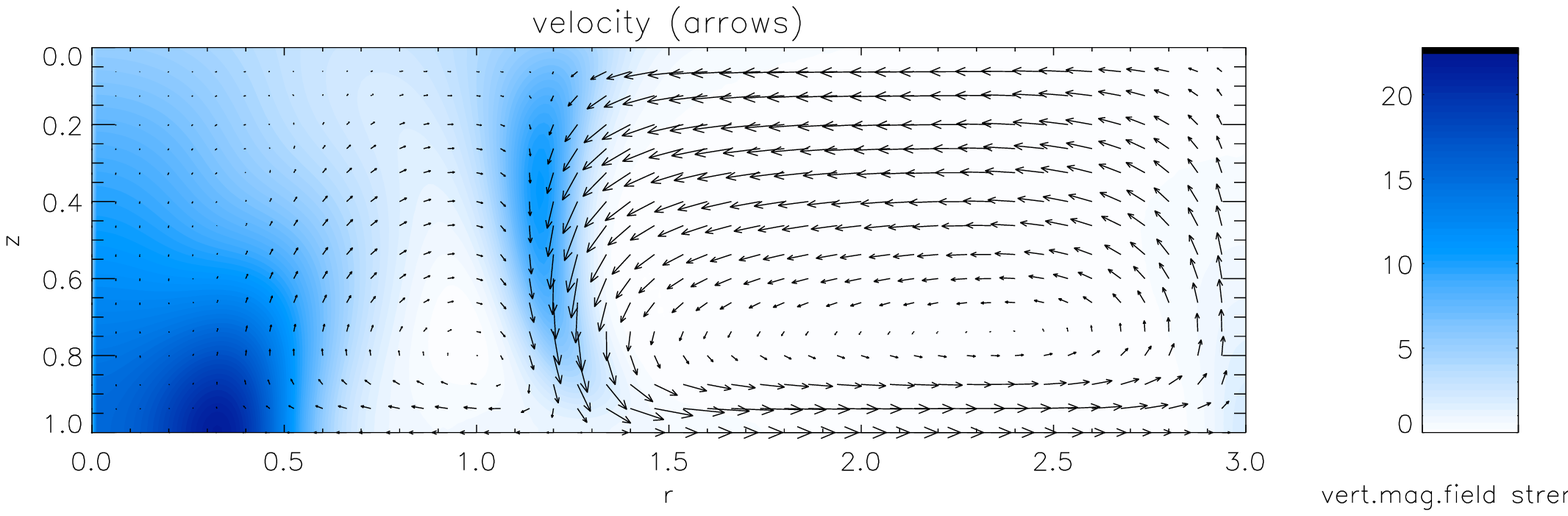}
	\vspace*{0.25cm}
	\caption{Dependence on the rotation rate. Results are shown for the same parameters as the reference case (figure~\ref{Fig:Results:Repro}) except the rotation rate, $\Omega$, is reduced here from 0.1 to 0.02 (left panel) and 0 (right panel), respectively. }
	\label{Fig:Results:WithRotationRate}
\end{figure}

This work is an extension of \opencite{2006MNRAS.369.1611B} and \opencite{2008MNRAS.387.1445B} in which we are exploring the conditions under which a stable flux concentration with an outflow over an inflow can exist.
Unless otherwise specified, the simulation parameters for the results presented here are: $Q=32$, $m=1$, $\Gamma=3$, $\zeta_0=0.2$, $\Omega=0.1$, $\sigma=1$, $\theta=10$, and $R=10^5$, referred to as the reference case in the text below.
This is one of the few cases in which a diverging over converging flow structure forms.
Simulation results for the same parameters were originally presented in figure~17 of  \opencite{2006MNRAS.369.1611B}.
A visualization of the flow and the magnetic field strength in shown here in figure~\ref{Fig:Results:Repro}.
It seems that externally applied rotation is crucially important in maintaining this flow structure.
Figure~\ref{Fig:Results:WithRotationRate} shows two examples in which we reduced the rotation rate resulting in a weakening of this flows configuration.
In fact, we have not found a parameter set without rotation in which a stable outflow over inflows was established that at the same time tightly held the flux concentration together.

\begin{figure}
	\centering
	\vspace*{0.25cm}
	\includegraphics[width=5.6cm]{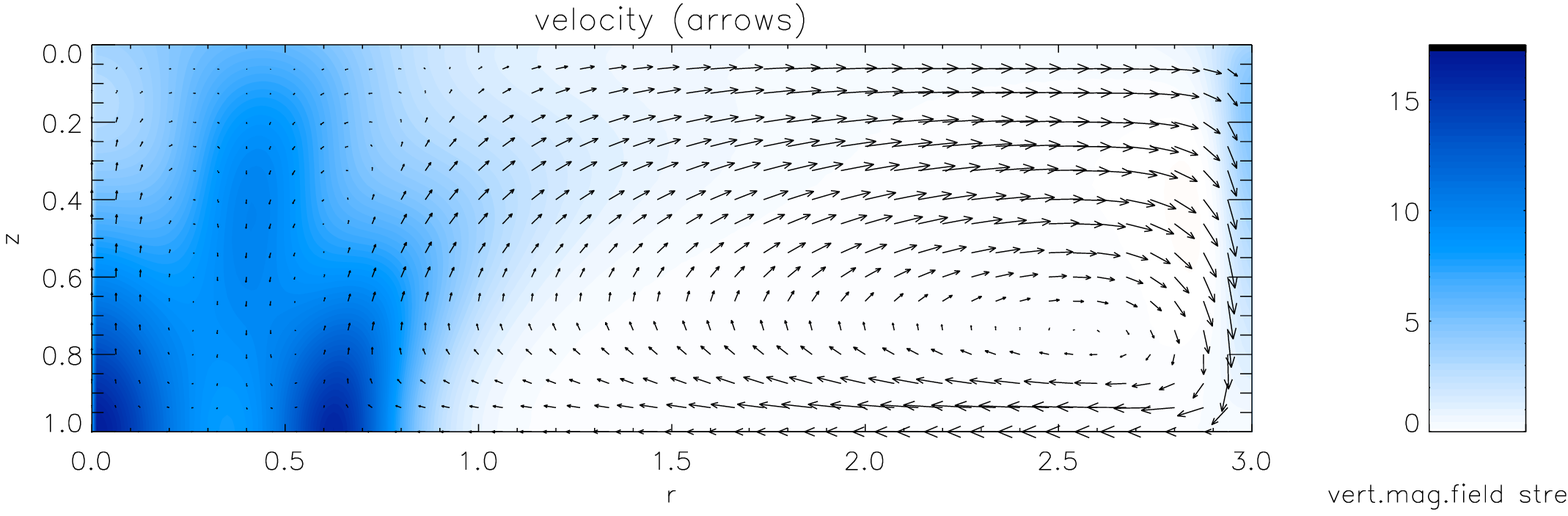} \hspace*{0.5cm}
	\includegraphics[width=5.6cm]{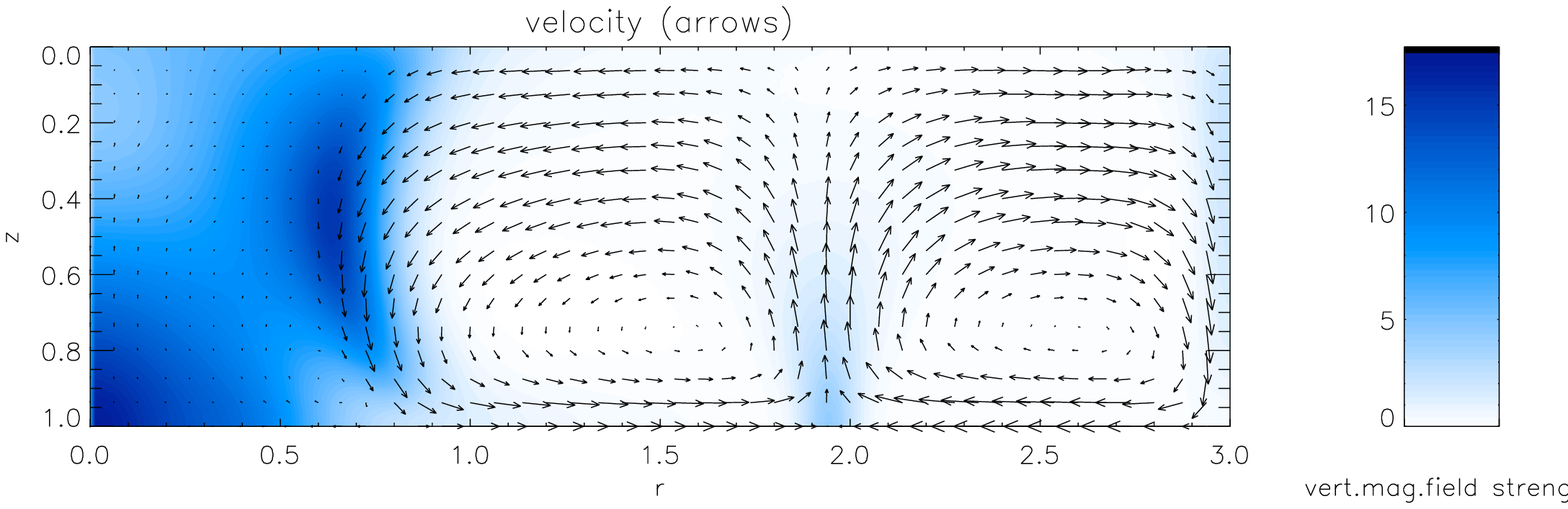}
	\vspace*{0.25cm}
	\caption{Prandtl number dependence. Results are shown for the same parameters as the reference case (figure~\ref{Fig:Results:Repro}) except the Prandtl number, $\sigma$, has been reduced from 1.0 to 0.3 (left panel) and 0.1 (right panel), respectively. }
	\label{Fig:Results:WithPrandtlNumber}
\end{figure}

Similarly, reducing the Prandtl number, $\sigma$, to more realistic values weakens this outflow configuration and a configuration with inflows over outflows, more similar to the observational evidence, is found.
This is illustrated in figure~\ref{Fig:Results:WithPrandtlNumber}.

\begin{figure}
	\centering
	\vspace*{0.25cm}
	\includegraphics[width=5.6cm]{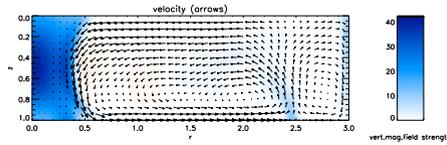}
	\vspace*{0.25cm}
	\caption{Same parameters as the reference case except for $\theta=20$ and $R=4\times10^5$.}
	\label{Fig:Results:WithR}
\end{figure}

Lastly, increasing the strength of the convection by increasing the Rayleigh number, $R$, also causes the flow to establish the more familiar converging over diverging flow as shown in figure~\ref{Fig:Results:WithR}. A weak secondary convective cell is formed in this case further away from the axis.

\section{Conclusions}

Admittedly, the simulations presented here are somewhat limited because of their simplifications and the restricted geometry, but they nevertheless can give us some insight into the  mechanisms that are involved in maintaining a tightly concentrated magnetic field near the solar surface inside structures such as magnetic pores or sunspots.
In fact, due to the simplicity of these simulations, a whole range of parameters can be explored rather easily.
The simulations have shown that for cases in which a stable, tightly concentrated magnetic structure is formed, most often a flow configuration with an inflow (towards the spot axis) above an outflow deeper below is established, compatible with observational evidence.
It is conceivable that such a configuration is better for maintaining the magnetic flux concentration at the center since the dynamic pressure of the inflow helps balancing the magnetic pressure inside the flux concentration. 
A flow with opposite sign is not impossible, though. 
In fact, under certain conditions a outflow configuration is established.
Rotation seems to be crucially important in these cases.

%

%

%
\bibliographystyle{spr-mp-sola}
\bibliography{hartlep}  

\end{article} 
\end{document}